\documentclass[iop]{emulateapj}

\newcommand{\eqref}[1]{(\ref{#1})}

\def\lesssim{\mathrel{\hbox{\rlap{\hbox{\lower4pt\hbox{$\sim$}}}\hbox{$<$}}}}
\def\gtrsim{\mathrel{\hbox{\rlap{\hbox{\lower4pt\hbox{$\sim$}}}\hbox{$>$}}}}

\shorttitle{SU Aurigae.}
\shortauthors{Akiyama et al.}

\usepackage{color}
\usepackage{floatrow}
\usepackage{ulem}
\usepackage{comment}
\begin{document}


\title{A TAIL STRUCTURE ASSOCIATED WITH PROTOPLANETARY DISK AROUND SU AURIGAE }


\author{EIJI AKIYAMA\altaffilmark{1}, EDUARD I. VOROBYOV\altaffilmark{2,3}, HAUYU BAOBABU LIU\altaffilmark{4,5}, RUOBING DONG\altaffilmark{5,6}, JEROME de LEON\altaffilmark{7}, SHENG-YUAN LIU\altaffilmark{5},  MOTOHIDE TAMURA\altaffilmark{7,8,9}}
\affil{
\altaffilmark{1}Institute for the Advancement of Higher Education, Hokkaido University, Kita17, Nishi8, Kita-ku, Sapporo, 060-0817, Japan; eakiyama@high.hokudai.ac.jp \\
\altaffilmark{2}Department of Astrophysics, University of Vienna, Vienna 1018, Austria \\
\altaffilmark{3}Research Institute of Physics, Southern Federal University, Rostov-on-Don, 344090, Russia \\
\altaffilmark{4}European Southern Observatory, Karl Schwarzschild Str 2, 85748 Garching bei M\"{u}nchen, Germany \\
\altaffilmark{5}Institute of Astronomy and Astrophysics, Academia Sinica, P.O. Box 23-141, Taipei, 10617, Taiwan \\
\altaffilmark{6}Department of Astronomy/Steward Observatory, The University of Arizona, 933 North Cherry Avenue, Tucson, AZ 85721, USA \\
\altaffilmark{7}Department of Astronomy, The University of Tokyo, 7-3-1, Hongo, Bunkyo-ku, Tokyo, 113-0033, Japan \\
\altaffilmark{8}Astrobiology Center of NINS, 2-21-1, Osawa, Mitaka, Tokyo, 181-8588, Japan \\
\altaffilmark{9}National Astronomical Observatory of Japan, 2-21-1, Osawa, Mitaka, Tokyo, 181-8588, Japan \\
}

\begin{abstract}
We present Atacama Large Millimeter/submillimeter Array (ALMA) observations of the CO ($J$=2--1) line emission from the protoplanetary disk around T-Tauri star SU Aurigae (hereafter SU Aur). Previous observations in optical and near infrared wavelengths find a unique structure in SU Aur. One of the highlights of the observational results is that an extended tail-like structure is associated with the disk, indicating mass transfer from or into the disk. Here we report the discovery of the counterpart of the tail-like structure in CO gas extending more than 1000 au long. Based on geometric and kinematic perspectives, both of the disk and the tail-like structure components physically connect to each other. Several theoretical studies predicted the observed tail-like structure via the following possible scenarios, 1) a gaseous stream from the molecular cloud remnant, 2) collision with a (sub)stellar intruder or a gaseous blob from the ambient cloud, and 3) ejection of a planetary or brown dwarf mass object due to gravitational instability via multi-body gravitational interaction. Since the tail-like structures associated with the SU Aur disk is a new example following RW Aurigae, some disks may experience the internal or external interaction and drastically lose mass during disk evolution.
\end{abstract}



\section{Introduction}
Recent observations by the Atacama Large Millimeter and Submillimeter Array (ALMA) revealed that there are a few of protoplanetary disks that have a tail-like structure or a tidal stream is associated with them in the Taurus-Auriga star forming region. It is considered as a signature that disk may be gravitationally influenced by internal or external agents and lose a large amount of disk mass during their evolution. Such events affect the development of planetary system and may produce a wide variety of planet configurations reported in exoplanet observations.

Previous observations in optical wavelength towards a protoplanetary disk around a young T-Tauri star SU Aurigae (hereafter SU Aur) suggested that reflection nebulae were associated with SU Aur. Such a reflection nebula usually seen in YSOs can be attributed to an outflow cavity as a result of stellar emission reflection at the cavity wall (e.g., \citealt{tamura91,lucas98a,lucas98b,pad99}). However, recent observations in optical and near infrared wavelengths suggested a different origin for SU Aur, and revealed a protoplanetary disk around SU Aur and an extended tail-like structure associated with the disk \citep{grady01,chak04,bertout06,jeffers14,leon15}. Possible explanations for the origin of the tail are discussed in \citet{leon15} and they suggest that tidal interaction with a brown dwarf is the the most plausible mechanism. As a result, the tail-like structure may be a consequence that the disk can obtain or lose some amount of mass via mass transfer from or onto the disk. 

In the standard model of planet formation, kilometer-sized objects, so-called planetesimals, are formed by coagulation of dust grains and planetesimals are responsible for forming more massive bodies, planetary embryos or protoplanets in a protoplanetary disk \citep{hayashi85}. If the tail-like structure is a tidal stream, a huge number of small mass objects including (proto)planets that resides in a protoplanetary disk known as a planet nursery must be released with gas and dust in the ambient, which might be a clue for the origin of the isolated planetary mass objects. In the observations by the IRAM Plateau de Bure Interferometer (PdBI) and ALMA, RW Aurigae system is also an established example in which a recent stellar flyby is believed to induce a tidal stream \citep{cabrit06,rodrigues18} and the detailed smooth particle hydrodynamics (SPH) simulations reproduced in detail the observations of the system \citep{dai15}.

As part of the Strategic Explorations of Exoplanets and Disks with Subaru (SEEDS) project \citep{tamura09}, we conducted $H$-band polarimetry observations toward protoplanetary disk around SU Aur in Taurus-Auriga star formation region which is at a distance of $158\pm2$ pc \citep{gaia18}. SU Aur is a $6.8\pm0.1$ million year old \citep{bertout07} G1 -- 2 type star \citep{ricci10,calvet04,herbig52}, whose stellar mass is $1.7\pm0.2 M_\odot$ \citep{bertout07,calvet04}, dust disk mass is $9\times10^{-4}M_\odot$ \citep{ricci10}, and mass accretion onto the central star is 0.5 -- 3 $\times$ 10$^{-8} M_\odot$ / year \citep{calvet04}. It is surrounded by a parental molecular cloud and its extinction ($A{\rm v}$) is 0.9 mag \citep{bertout07}. The $H$-band scattered light observation found that small dust grains, typically 0.1 -- 1 $\mu$m size, form a long tail-like structure of more than 350 au in length extending from the disk body to the west \citep{leon15}. Soon after the Subaru observation, CO observations by ALMA revealed the same tail-like structure more than 1000 au long extending from the northern part of the disk body to the southwest direction. 

In this paper, we describe the details of the CO ($J$=2--1) ALMA observations and data reduction in section \ref{obs}. In section \ref{result}, the details of morphological and kinematical structures of disk and tail-like structure are shown. Several new findings by observations are also provided. In section \ref{discussion}, possible mechanisms for generating tail-like structure are discussed based on several theoretical works in the last decade. Finally, we discussed the origin of the tail-like structure. Section \ref{summary} summarizes our findings and implications. 

\section{Observations and Data Reduction}
\label{obs}
The ALMA archival data (2013.1.00426.S) are used in the paper. Observations were performed on July 19 and August 8 in 2015 at Band 6, containing 36 and 40 12-m antennas, respectively. The baseline length of the observed array configuration spanned from 15 m to 1.57 km, resulting in a maximum recoverable size of 11 arcsec. J0423-0120 was used as a bandpass calibrator and J0510+1800 (J0510+180) are used to calibrate for amplitude and phase. The 2 SB receivers were tuned to observe the CO ($J$=2--1) rest frequency of 230.538 GHz and the nearby continuum at 1.3 mm in wavelength. The spectral resolution was set up to be 30.518 kHz, corresponding to 0.04 km s$^{-1}$. The average precipitable water vapor (PWV) during observation was about 1.38 mm. The Band 6 data were imaged using the CLEAN task provided in CASA version 4.3.1 package, and self-calibration was applied to correct for short-term phase variation in the continuum data. We checked rms noise level at each round by generating the image and stopped self-calibration with amplitude and phase correction when the rms noise level reached minimum. Then we apply the calibration table to the CO ($J$=2--1) line data and generate the images with a robust parameter of 1.5 in Briggs weighting. As a result of all calibrations, the synthesized beam of 0."235 $\times$ 0."356 is obtained.

Finally, the sensitivity of the primary beam is scaled by primary beam correction task so that the sensitivity becomes uniform in the entire field of view (FoV). Note that the sensitivity of the FoV boundary generally becomes quite low in comparison with the one at the FoV center. The emission becomes noisy and blurred along the boundary. The primary beam correction task increases the emission at the FoV boundary by simply re-scaling the sensitivity in the entire FoV to equate with the sensitivity level at the FoV center, and noise is also simultaneously increases by the same amount. Therefore, the signal-to-noise (S/N) ratio still remains the same.

\section{Results}
\label{result}
The moment 0 (integrated intensity) map in CO ($J$=2--1) of the protoplanetary disk around SU Aur is presented in Figure \ref{fig1}a. The figure shows a disk component and a tail-like structure more than 1000 au long extending from the northern part of the disk body to the southwest direction. The tail-like structure of small dust grains that extends at least 350 au in the same orientation was also observed in the $H$-band by Subaru telescope \citep{leon15}. Thus, the tail-like structure in gas significantly extends beyond the structure in 0.1 -- 1 $\mu$m sized dust grain. The disk component has a peak intensity of 400 mJy beam$^{-1}$ km s$^{-1}$ in CO ($J$=2--1) emission line. The resultant rms is 28 mJy beam$^{-1}$ km s$^{-1}$ in the integrated time on source of 14.5 minutes after data flagging, providing an S/N ratio of approximately 14. Note that the fan-shaped emission seen in southwest region in the moment maps could possibly be artificial. The sensitivity of the primary beam decreases as a function of gaussian distribution, and thus the sensitivity at the FoV boundary is a half of the value at the FoV center in ALMA\footnote{The antenna power response data and the detail description are provided in Chapter 3 of the ALMA Cycle 6 Technical Handbook.}. Since the tail-like structure reaches very close to the FoV circular boundary, just outside of the imaging area, the emission may spread out in a fan-shaped form in the southwest direction due to smearing effect.

No signature of collimated jet, outflow, or outflow cavity that previously reported in the optical and IR observations \citep{nakajima95,chak04} is detected in the CO and mm continuum observations at least within a 15 arcsec region from a central star. Figure \ref{fig1}b displays the moment 1 (velocity distribution) map. It shows disk rotation and tail-like component with its $V_{\rm LSR}$ at roughly 6.5 km s$^{-1}$ uniformly extending to the southwest direction. The velocity gradient is twisted from SE -- NW direction in the very inner part to NE -- SW in the outer part, indicating that the disk could be twisted or warped although the possibility of infall around the disk can not be fully discarded \citep{facchini18}. Figure \ref{fig1}c represents a moment 2 (velocity dispersion) map. It also shows both disk and tail-like components with similar morphology with the moment 0 and 1 maps. An interesting feature in the moment 2 map is the high velocity dispersion region displaying in a boomerang shape near the center of the disk. It extends along the tail-like structure then disappears down the tail. The reflection nebulae that emission scattered at the outflow cavity wall ($ref.$ \citet{tamura91,fuente98c,pad99}) previously reported is probably a part of tail-like structure obtained in CO ALMA observation. 

In order to directly compare dust distribution with CO, we provide a superposed image of $H$-band and 1.3 mm continuum emission that traces 0.1 - 1 $\mu$m- and mm-sized dust, respectively, on CO map as shown in Figure \ref{fig2}. Since the disk component and its boundary are not clear in the intensity map, the moment 1 (velocity distribution) map that shows the disk component more clearly by rotational motion is overlaid for reference purpose. Unlike CO tail-like feature, compact structure is detected in 1.3 mm continuum emission, whose peak intensity of 850 mJy beam$^{-1}$, as displayed in white contour. $H$-band emission, on the other hand, shows a tail-like structure in the same orientation with the one seen CO emission. The results support that 0.1 -- 1 $\mu$m-sized dust is well coupled with gas, while mm-sized dust is not. It should be noted that the possible reason of non-detection of mm-sized dust in the tail-like structure could be sensitivity limit since the dust mass in the tail-like structure is estimated to be $\sim 6\times10^{-8}M_\odot$ \citep{leon15}, too small amount of dust to detect in the observation. The tail-like structure in mm sized dust can not be discarded at this moment.

We generate a velocity profile of the disk component that covers the region of a radius of 1.2 arcsec, corresponding to $\sim$ 172 au, from the central star (Figure \ref{fig3}a). It shows a doubled peaked profile separated into the blueshift (gas approaching) and redshift (gas receding) components at {5.8 $\pm$ 0.3 km s$^{-1}$. It is of high interest that the red-shifted emission is more dominant than the blue-shifted component, indicating that some additional velocity component is associated with the disk other than disk rotation. Figure \ref{fig3}b and c show profiles of the velocity and velocity dispersion curves along the tail-like structure shown in Figure \ref{fig1}b and c, respectively. A point of the origin on the distance is indicated by a cross mark in the figures. It is noteworthy that the velocity structure provided by 1st and 2nd moment maps (Figure \ref{fig1}b and c) shows the disk rotation smoothly varied to linear motion forming into the collimated tail-like structure and maintains almost constant velocity of $\sim$ 6.5 km s$^{-1}$ without any discontinuity. Thus, the tail-like component is red shifted by 0.5 -- 1 km s$^{-1}$ from the systemic velocity of 5.8 km s$^{-1}$. The tail component has a velocity dispersion of $\sim$ 0.7 km s$^{-1}$ whereas the disk component has $\sim$ 2 km s$^{-1}$. 

Figure \ref{fig4} shows a velocity channel map with $\sim$ 0.3 km s$^{-1}$ intervals for the CO ($J$=2--1) emission line. The interesting tail-like feature of importance is detected from 5.14 to 8.16 km s$^{-1}$ and most of the tail-like structure is red-shifted from the systemic velocity of 5.8 km s$^{-1}$. It is even more noteworthy that the tail-like structure can be seen as two components, a long tail-like structure more than 1000 au extending to the west seen between 5.14 and 6.94 km s$^{-1}$ and a relatively small elongated structure with a few 100 au long extending to the (south)east direction between 7.25 and 8.16 km s$^{-1}$. The detected two components are morphologically consistent with our $H$-band result previously observed by the Subaru telescope (see Figure 1 and 3 in \citealt{leon15}).

\section{Discussion} 
\label{discussion}
\subsection{Kinematic Structure of Star and Disk System} 
\label{kinematic}
The moment maps and several channels in the velocity channel map reveal that SU Aur has a distinctive tail-like structure that is kinematically connected with disk component. It is beneficial to see non-keplerian motion in the disk+tail system for discussing the velocity structure and its origin. To investigate the interaction between the disk and the tail-like structure, it is important to differentiate their motions and analyze each component individually.

We made a simple keplerian disk model in the same orientation with the observed disk component and then computed the residual after subtracting the model from the observed moment 1 map. In the model, the disk is treated as gas moving in the keplerian velocity along the circular orbit around the central star. Detail description can be found in the Appendix in \citep{isella07}. We also assume the disk has flat geometry in the model. Figure \ref{fig5}a shows the same velocity distribution map with Figure \ref{fig1}b. Figure \ref{fig5}b shows the residual after subtracting the keplerian disk model shown in Figure \ref{fig5}c from the observed velocity distribution shown in Figure \ref{fig5}a. Note that we focus on the deviation of the disk rotation and subtraction applied only to the disk component. Thus, the velocity of tail-like component remains the same as original. As a result of subtraction, the motion in the southern part of the disk is almost cancelled by the model counterpart, resulting in approximately zero velocity. Therefore, it supports that the keplerian rotation in the southern part of the disk remains almost intact. On the other hand, the velocity component is not sufficiently subtracted in the northern part of the disk and still remains some velocity component by 4 km s$^{-1}$ from the systemic velocity. It implies that gas in the northern part of the disk dose not have keprarian rotation and the red-shift component dominates in the region. The result is consistent with the observationally derived velocity profile of the disk given in Figure \ref{fig3}a. 

The detected tail-like structure that is connected with the northern part of the disk can also be traced by the channel map shown in Figure \ref{fig5}. Since the motion of the tail-like structure is smoothly connected with only northern part of the disk, it must be responsible for the velocity deviation from keplerian motion as shown in the residual mentioned in the above. Another feature of interest is that the tail-like structure has two velocity components. The channel map shows that a red-shift component between 5.14 to 6.94 km s$^{-1}$ extends more than 1000 au to the west from the disk, while the other seen between 7.25 and 8.16 km s$^{-1}$ extends a few hundred AU to the east from the disk. Similar structure is seen in RW Aur. binary or triple star system \citep{ghez93} as tidal sculpting probably caused by gravitational interaction due to stellar encounter. They indicate that the tidal streams should be a result of stripping the primary protoplanetary disk around RW Aur A by the fly-by of a companion RW Aur B \citep{cabrit06,rodrigues18}. The fact that keplerian and non-keplerian motion simultaneously exist in the disk implies that the disk must be influenced by some kind of internal or external agent that significantly destroys keplerian rotation. We discuss several possible mechanisms that may have impacted the SU Aur system in the following section.   

\subsection{Origin of the Tail-Like Structure}
\label{origin}
Several mechanisms can theoretically explain the observed tail-like structure. It can be either a result of close encounter with a (sub)stellar object, or ejection of a small object or gaseous clumps formed in the disk via gravitational fragmentation, or even an interaction with the external parental cloud. We here discuss the possible scenarios for the formation of the tail-like structure based on recent theoretical works.

Firstly, the observed extended tail-like structure can be explained by a close encounter with a (sub)stellar object passing through the outer regions of the disk. The collider drags some disk material and makes one side of the disk elongated to the direction of the collider's trajectory when it passes close to the outer disk edge. We note that the tail becomes curved with time, perhaps  as a response to disk rotation. Such a picture is demonstrated in smoothed particle hydrodynamics (SPH) simulations of \citet{forgan09} and \citet{thies11}, showing that the encounter strips the outer regions of the disk through tidal tails. \citet{voro17} carefully investigate the results after setting a hypothetical intruder 3000 au away from the disk and sending towards the target either on a prograde or retrograde trajectory. As results, a long tail and its counter part are generated after collision. The SPH simulation in RW Aur system is an established case in which a recent stellar flyby is believed to induce a major streamer \citep{dai15}. A possibility of the disk-disk encounter is also reported in HV and DO Tau \citep{winter18}. HV Tau triple system initially formed a quadruple system with DO Tau, and then the system experienced an encounter among themselves due to the highly eccentric orbit of DO Tau. The SPH simulation shows that a pair of the observed tidal tails can be induced by the internal encounter in the same system and subsequent disk evolution. All of the simulations indicate that gravitational interaction between an intruder and a target generates a tail-like structure and its counter part. 

Although an encounter between stars is rare even in populous regions such as Orion nebula cluster, the situation would be different for disks. According to numerical N-body simulations, approximately 8 \% of disks with a radius of $\sim$ 300 au like SU Aur disk will experience at least one encounter in 2.89 Myr \citep{scally01}. The events of encounters between (sub)stellar objects and disks are thought to be much more likely than those of star by star collision. Moreover, close stellar encounters with disks are often invoked to explain the seemingly compact disk configurations as compared to what is usually obtained for the model disks formed in isolation. 

If this is the case, the intruder likely moves from the east to the west direction and gradually spread the material out and make a fan-like form as it approaches to the FoV boundary. Since the tail-like component is red shifted and no sudden change of velocity is seen in the velocity distribution curve as shown in Figure \ref{fig3}b, it seems that the intruder approaches from the near side of the disk from the observer to the other side. It is known from 2MASS, Spitzer, and WISE surveys\footnote{The data is open for public use and can be obtained from Aladin Sky Atlas, https://aladin.u-strasbg.fr.} that quite a number of point sources densely populate the Auriga region and SU Aur is surrounded by them. Stellar encounter could occur in any direction.

However, it could be opposite direction if we see gas in the streamer is moving towards the star. In that case, the intruder should be a gaseous blob, not a (sub)stellar point source. The gaseous blob would not generate a tail-like structure, rather it  would break into several parts by tidal force until it hits the disk. Although no numerical model that specifically shows gaseous blob collision is reported, our kinematic analysis does not exclude the possibility of a gaseous blob collision because the velocity distribution curve shows gradual acceleration from a distance of 10 arcsec toward the center. Furthermore, from the standpoint of energy conservation, the degree of red shift is consistent when the velocity estimated by the kinetic energy that is converted from the potential energy initially retained before the blob approaches to the disk. If this is the case, the gaseous blob is stretched as it approaches to the disk with acceleration, resulting in a tail-like structure. 

The boomerang shape in the disk component seen in the velocity dispersion map shown in Figure \ref{fig1}c also supports a scenario of (sub)stellar encounter or a gaseous blob collision onto the disk. A curl of motion of gas changes into a linear motion along the tail direction rather than random motion and velocity dispersion disappears down the tail. It implies that northern part of the disk gas is influenced by either pushing or dragging force externally acting on the disk along southwest or northeast direction. We also estimated the brightness temperature of the tail-like structure and check whether or not the tail-like structure is optically thick or not. We selected a panel detecting tail-like structure from the channel map and applied the beam size and the rest frequency in CO ($J$=2--1) line. As a result of estimation, the brightness temperature is about 16 K, indicating the CO emission is optically thick because the temperature saturates to 10 - 20 K. The tail-like component should be located in front of the disk because northern part of the disk is partially hidden behind the sufficiently optically thick CO tail-like structure as shown in Figure \ref{fig1}b. The disk might have a flare-up structure because the far-side of the disk can face toward the observer with larger surface that make the red shifted component have larger apparent area than the blue-shifted counterpart. 

Secondly, based on the moment 1 map, there is a possibility that the disk interacts with the external molecular cloud, which channels its material to the disk via a filamentary-like stream. This situation can be seen in numerical simulations of a pre-stellar core collapse and disk formation, which self-consistently follow the core to star plus disk transition and relax the assumption of the fixed central star, i.e., allow for stellar motion in response to the gravitational potential of the disk plus envelope system \citep{rv17}. 
Since SU Aur does not sit directly in the center of the cloud gravitational potential well, the matter of the cloud first falls towards its global potential well and then is attracted towards SU Aur in the form of a narrow stream.

In fact, SU Aur is immediately adjacent to a more massive Herbig Ae star, AB Aurigae (hereafter AB Aur) with $2.4\pm0.2 M_\odot$ \citep{dewarf03}. It is 3 arcmin ($\sim$ 26000 au) away to the west from SU Aur and the tail-like structure points toward AB Aur. A massive molecular cloud surround them and mass flow emerges in a direction dominantly to the AB Aur and more material would likely be supplied to AB Aur rather than SU Aur. Therefore, SU Aur can experience mass supply from only one direction due to disproportionate flow, which results in a tail-like structure. Note that the tail-like structure is too extended and reaches to the field of view (FoV) boundary of the observation, and thus the structure becomes noisy in the west part.

Finally, the tail-like structure can be generated by ejection of a planetary or brown dwarf mass object from the disk due to gravitational instability. This is invoked through growing random motion as several bodies gravitationally interacts in the disk \citep{basu12,voro15,voro16}. In this case, the disk should be massive enough to induce gravitational instability and disk fragmentation \citep{dunham14,dong16,liu16}. According to the millimeter observation of the PdBI, the disk mass can be $9.0\times10^{-2}M_\odot$ under assumption of gas to dust ratio of 100 \citep{ricci10}. This value is likely a lower limit, as the uncertainties in disk opacity, temperature, and  inclination could lead to disk mass underestimates by up to a factor of several \citep{dunham14}. Moreover, the ejected clumps can carry out a notable fraction of the original disk mass. According to \citet{voro13}, the minimum disk mass needed for clump ejection to operate is 0.1~$M_\odot$. Taken all these arguments together, the disk mass seem to be sufficient for the ejection scenario to occur in SU Aur. We note that the clump leaves behind a characteristic bow-shock tail as it propagates outward through the ambient medium.

In several preceding studies, several simulations reproduce the tail-like structure by all of the three above mechanisms. However, there are many things remains uncertain at this moment. For example, the steady mass flow in the molecular cloud is still hypothetical and the driving source is not observationally identified yet. The ejection scenario is possible but it has problems in subsequent results. Although the long tail structure seen in the observation is induced when a planetary mass object is ejected out of the disk, the disk must have more complex motion by interaction among the other remaining objects. Since no random motion is detected in the disk but rather disk motion uniformly connects with the motion of the tail-like structure, it seems less probable. On the other hand, collision with a (sub)stellar intruder or a gaseous blob falling onto the disk can explain more comfortably the generation of the tail-like structure. Kinematic structure in both of the disk and tail as seen in moment maps supports that SU Aur has a well-ordered velocity structure, and thus SU Aur probably experienced gravitational interaction by a single external force in the past. Since the upper limit of the mass of the intruder should be smaller than 2 $\times$ 10$^{-3} M_\odot$ based on the detection limit of the Subaru observation and no objects with 2 $\times$ 10$^{-3} M_\odot$ have been found around SU Aur, a gaseous blob accreting onto the disk is more probable explanation for generating the tail-like structure. In the event of the collision, some (proto)planets and planetesimals reside in the disk must be blown away, resulting in free-floating planets. Note that the possibility of a (sub)stellar encounter can not be fully excluded because uncertainty still remains in estimating object mass and our FoV is limited. The observational data should be improved by follow-up observations with larger FoV and high sensitivity, and we leave new observations including corroborative simulations for SU Aur for our future work.

\section{Summary}
\label{summary}
As results of ALMA millimeter observations, characteristic appearance of SU Aur has become clear and a tail-like structure is associated with the disk around SU Aur. The geometric-kinematic analysis of the observed tail-like structure reveals the positional relation of disk+tail system and mass flow motion along the tail-like structure. We consider three possible scenarios; 1) collision with a (sub)stellar intruder or a gaseous blob from the ambient cloud, 2) interaction with the external cloud channeling its material towards the disk in a filamentary-like stream, and 3) ejection of a planetary or brown dwarf mass object due to gravitational instability via multi-body gravitational interaction. Among of the three possibilities, the collision with a (sub)stellar intruder or a gaseous blob is probably the most plausible explanation based on physical properties of SU Aur. Follow-up observation by shock-tracer such as SO emission line would be productive for confining the origin of the tail-like structure. 

\acknowledgments
E. A. is supported by MEXT/JSPS KAKENHI grant No. 17K05399. M. T. is supported by MEXT/JSPS KAKENHI grant Nos. 18H05442, 15H02063, and 22000005. E. Vorobyov acknowledges financial support from the Russian Foundation for Basic Research (RFBR), Russian-Taiwanese project \#19-52-52011. H.B.L. is supported by the Ministry of Science and Technology (MoST) of Taiwan (Grant Nos. 108-2112-M-001-002-MY3 and 108-2923-M-001-006-MY3). This paper makes use of the following ALMA data: ADS/JAO.ALMA\#2013.1.00426.S. ALMA is a partnership of ESO (representing its member states), NSF (USA) and NINS (Japan), together with NRC (Canada), MOST and ASIAA (Taiwan), and KASI (Republic of Korea), in cooperation with the Republic of Chile. The Joint ALMA Observatory is operated by ESO, AUI/NRAO, and NAOJ. The National Radio Astronomy Observatory is a facility of the National Science Foundation operated under cooperative agreement by Associated Universities, Inc. Data analysis was carried out on the Multi-wavelength Data Analysis System operated by the Astronomy Data Center (ADC), National Astronomical Observatory of Japan. We appreciate Munetake Momose for technical advice in data analysis. We would like to thank the anonymous referee who helped improve the manuscript.

\clearpage
\begin{figure}[t]
\begin{center}
\includegraphics[scale=0.6]{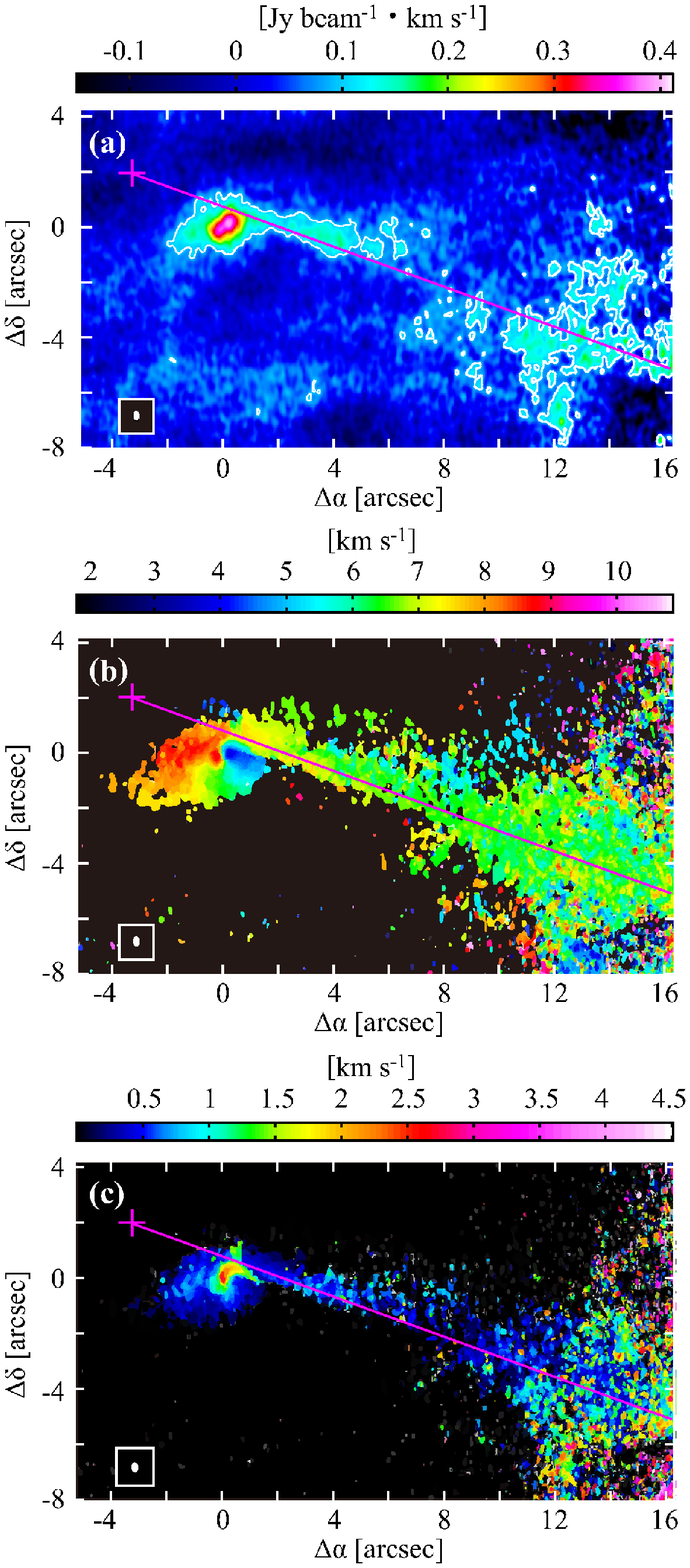}
\caption{The 0th (integrated intensity), 1st (velocity distribution), and 2nd (velocity dispersion) moment maps of CO($J$ = 2 -- 1) emission are provided in panels (a), (b), and (c), respectively. A contour in the panel (a) denotes the emission more than 3 $\sigma$ (1 $\sigma$ = 23 mJy beam$^{-1}$). In the panel (b), the gas velocity is given in color gradient: blue (gas approaching) and red (gas receding). The CO($J$ = 2 -- 1) moment maps are generated from the archive data (2013.1.00426.S). The synthesized beam is given by a filled ellipse at the lower left corner in each panel. The red solid line shown in panels (b) and (c) represents the direction that is used in velocity and velocity dispersion curves provided in Figure \ref{fig3}b and c, respectively. They are similarly distributed but not exactly coincide each other because the tail-like structure seen in both of the maps do not exactly matches each other. The 1st and 2nd moment maps is created using velocity channels from 1.87 to 10.87 km s$^{-1}$ and emission above 30 mJy beam$^{-1}$. The cross mark (R.A. 04h55m59$\arcsec$.73, Dec. 30$\arcdeg$34$\arcmin$03$\arcsec$.32) in each panel indicates the zero point on the distance used in velocity distribution and velocity dispersion curves shown in the panels (b) and (c) of Figure \ref{fig3}. North is up and east is to the left in all of the panels. Note that the circular boundary of the observation FoV is just outside of the image (not visible int the Figure {fig1}) and the tail-like structure reaches to it.}
\label{fig1}
\end{center}
\end{figure}

\begin{figure}
\begin{center}
\centering 
\includegraphics[scale=0.6]{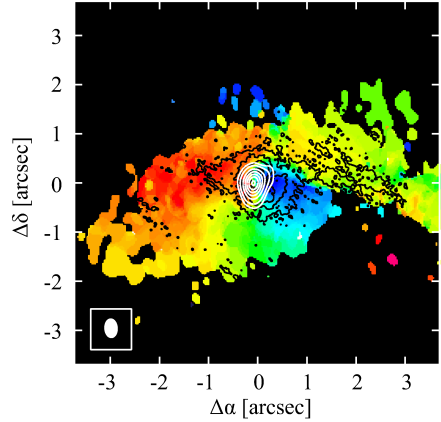}
\caption{CO($J$ = 2 -- 1) moment 2 map of the disk component is given in a larger scale of Figure \ref{fig1}a with Subaru $H$-band polarized intensity (PI) \citep{leon15} in dark color contour and ALMA 1.3 mm continuum in white contour, respectively. The dark colored contour of PI represents the region more than 2 $\sigma$ emission (more detail image is available in Figure 1 in \citep{leon15}). In the 1.3 mm continuum, the contours are 0.135 mJy beam$^{-1}$ (1 $\sigma$) $\times$ [5, 10, 20, 30, 40, 50, 60]. The color scale is the same as the one shown in Figure \ref{fig1}a.}
\label{fig2}
\end{center}
\end{figure}

\begin{figure}
\begin{center}
\centering 
\includegraphics[scale=0.37]{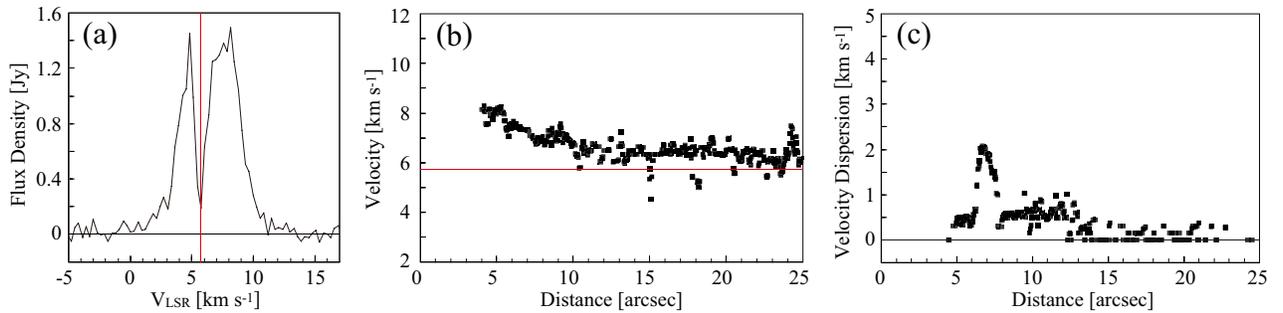}
\caption{Details of velocity structure of the disk and tail-like structures are given. Panel (a) represents a velocity profile of the disk component, the region of a radius of 1.2 arcsec, corresponding to $\sim$ 172 au from the central star. Panel (b) represents the velocity distribution curve along the tail-like structure as denoted by the red line in Figure \ref{fig1}b. The systemic velocity of 5.8 km s$^{-1}$ is shown by the red line in both panel (a) and (b). Panel (c) is the same as panel (b) for velocity dispersion. }
\label{fig3}
\end{center}
\end{figure}

\begin{figure}
\begin{center}
\centering 
\includegraphics[scale=0.75]{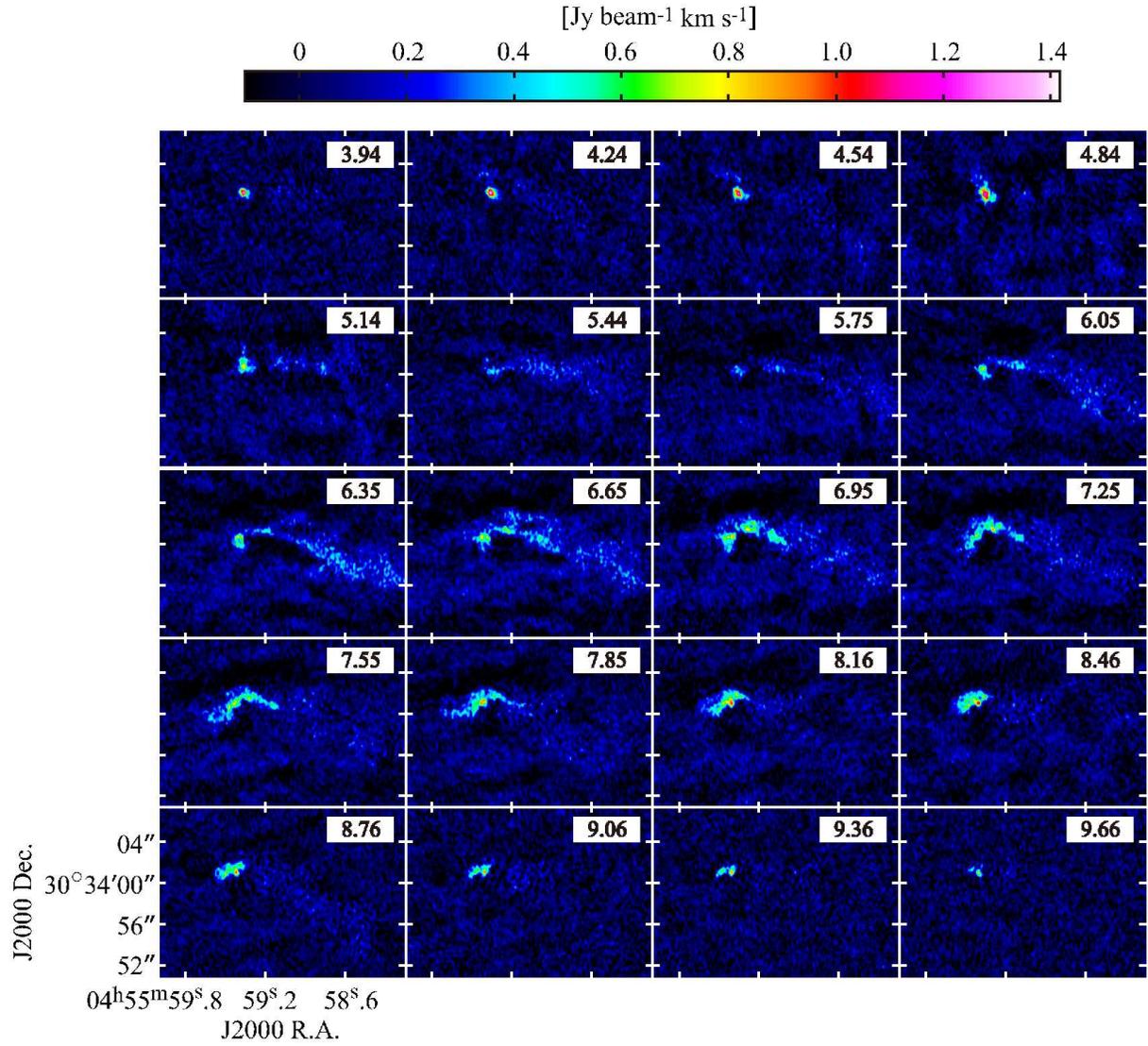}
\caption{Channel map with $\sim$ 0.3 km s$^{-1}$ velocity bins of the CO ($J$=2--1) line emission from the disk and tail-like structure associated with SU Aur. The LSR velocity in km s$^{-1}$ is provided in the upper right corner of each channel.}
\label{fig4}
\end{center} 
\end{figure}

\begin{figure}
\begin{center}
\centering 
\includegraphics[scale=0.6]{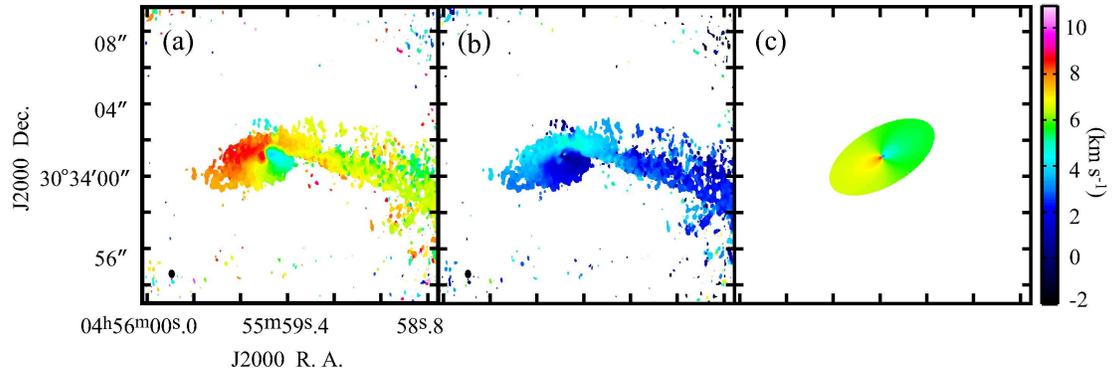}
\caption{Disk velocity structure analyses by model fitting is given. Panel (a) is the observed velocity distribution (moment 1) map of the SU Aur, a view of disk part shown in Figure \ref{fig1}b. Panel (b) is the residual after subtracting keplerian model shown in the panel (c) from observed velocity distribution in Figure \ref{fig3}a. The systemic velocity of 5.8 km s$^{-1}$ is given to the velocity component outside of the disk region so that the boundary between the disk and the tail-like structure is smoothly connected. Panel (c) is a simulated disk with keplerian rotation. Note that the systemic velocity of the SU Aur is shown by green color. See also Figure 6 in \citet{chak04} for the disk orientation based on HST/STIS observation.}
\label{fig5}
\end{center}
\end{figure}

\clearpage
\end{document}